\documentclass[showpacs,amsmath,amssymb,prb,twocolumn,superscriptaddress]{revtex4}
\usepackage[T1]{fontenc}
\usepackage{times}
\usepackage{graphicx}
\usepackage{wasysym}
\usepackage{subfigure}
\usepackage{amssymb,amsmath}
\usepackage{bm}

\begin{document}

\title{Optical Anisotropic Metamaterials: Negative refraction and focusing}

\author{Anan Fang}%

\affiliation{%
Department of Physics and Astronomy and Ames Laboratory,
Iowa State University, Ames, Iowa 50011, USA
}%

\author{Thomas Koschny}%
\affiliation{%
Department of Physics and Astronomy and Ames Laboratory,
Iowa State University, Ames, Iowa 50011, USA
}%
\affiliation{%
Institute of Electronic Structure and Laser, FORTH, Heraklion, 71110 Crete, Greece 
and Department of Materials Science and Technology, University of Crete, Heraklion, 71110 Crete, Greece
}%

\author{Costas M. Soukoulis}%
\email{soukoulis@ameslab.gov}
\affiliation{%
Department of Physics and Astronomy and Ames Laboratory,
Iowa State University, Ames, Iowa 50011, USA
}%
\affiliation{%
Institute of Electronic Structure and Laser, FORTH, Heraklion, 71110 Crete, Greece 
and Department of Materials Science and Technology, University of Crete, Heraklion, 71110 Crete, Greece
}%

\date{\today}

\begin{abstract}
We design 3D metallic nanowire media with different structures and numerically demonstrate they can be homogeneous effective indefinite anisotropic media by showing their dispersion relations are hyperbolic. For a finite slab, a nice fitting procedure is exploited to obtain the dispersion relations, from which we retrieve the effective permittivities. The pseudo focusing for the real 3D wire medium agrees very well with the homogeneous medium having the effective permittivity tensor of the wire medium. Studies also show that in the long wavelength limit, the hyperbolic dispersion relation of the 3D wire medium can be valid even for evanescent modes.
\end{abstract}
%


\maketitle

\section{Introduction \label {introduction}}

Recently, negative index materials (NIMs) and photonic crystals (PCs) are receiving more and more attention because of their extraordinary optical properties such as near field focusing, subwavelength imaging, and negative refraction. \cite {veselago68, pendry00, pendry06, notomi00, luo03, cubukcu03, foteinopoulou03, foteinopoulou05, moussa05, berrier04, schonbrun07, hu04} As first proposed, these NIMs have the permittivity, $\varepsilon$, and the permeability, $\mu$, simultaneously negative, which are achieved by overlapping electric and magnetic resonances. But the double resonance scheme also causes large resonance losses and technical difficulties in design and fabrication. In addition to negative index materials, both theoretical and experimental studies show the properties of negative refraction and subwavelength imaging can also occur in some uniaxially anisotropic media, which can have lower losses and be easier to fabricate. \cite {smith03,smith04, hoffman07, Silveirinha07, wangberg06, yao08, lindell01, scalora07, wood06, jacob06, Liu08} 

\begin{figure}
  \subfigure[]{
   \includegraphics[width=0.27\textwidth]{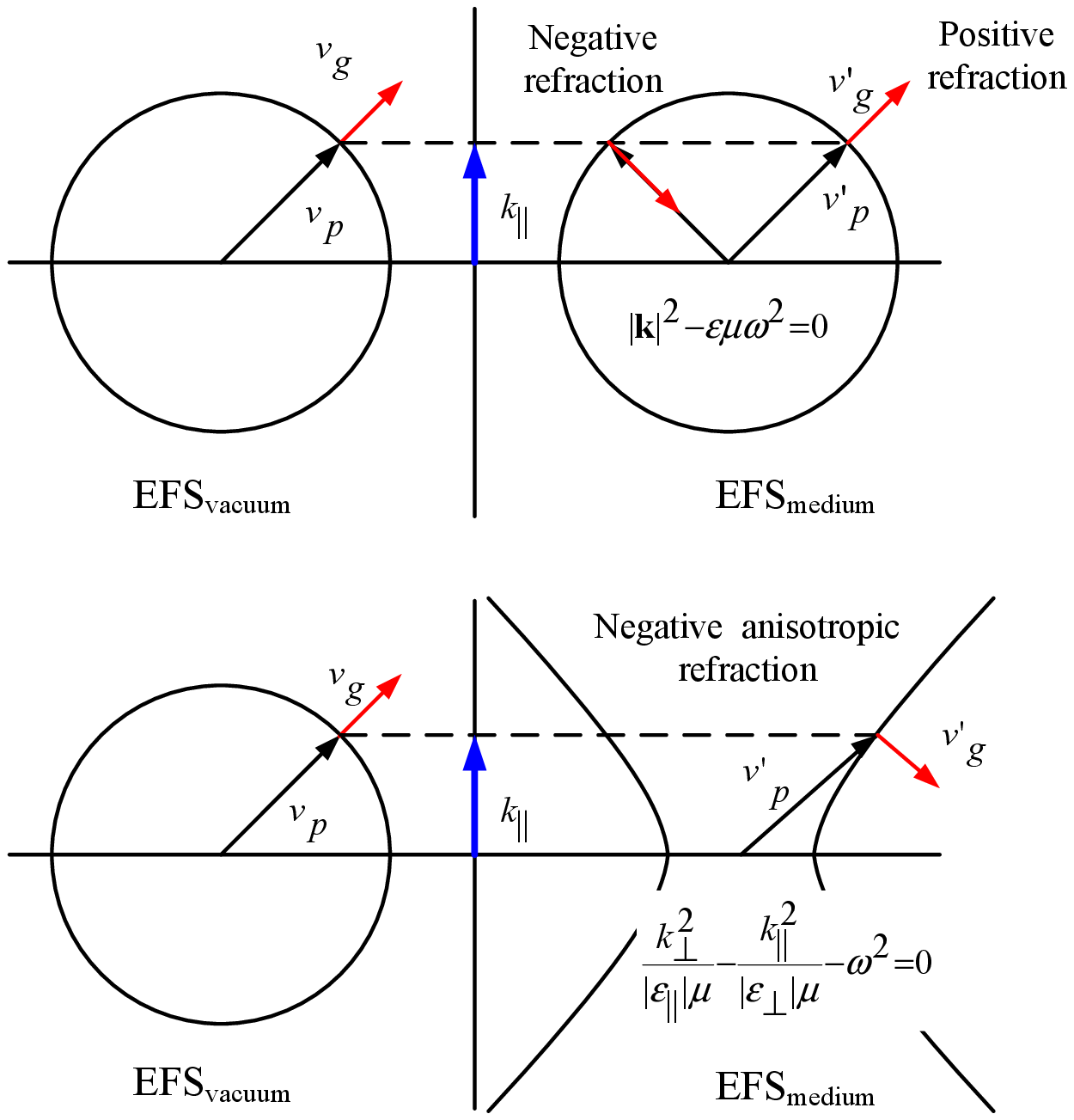}} 
  \subfigure[]{
   \includegraphics[width=0.18\textwidth]{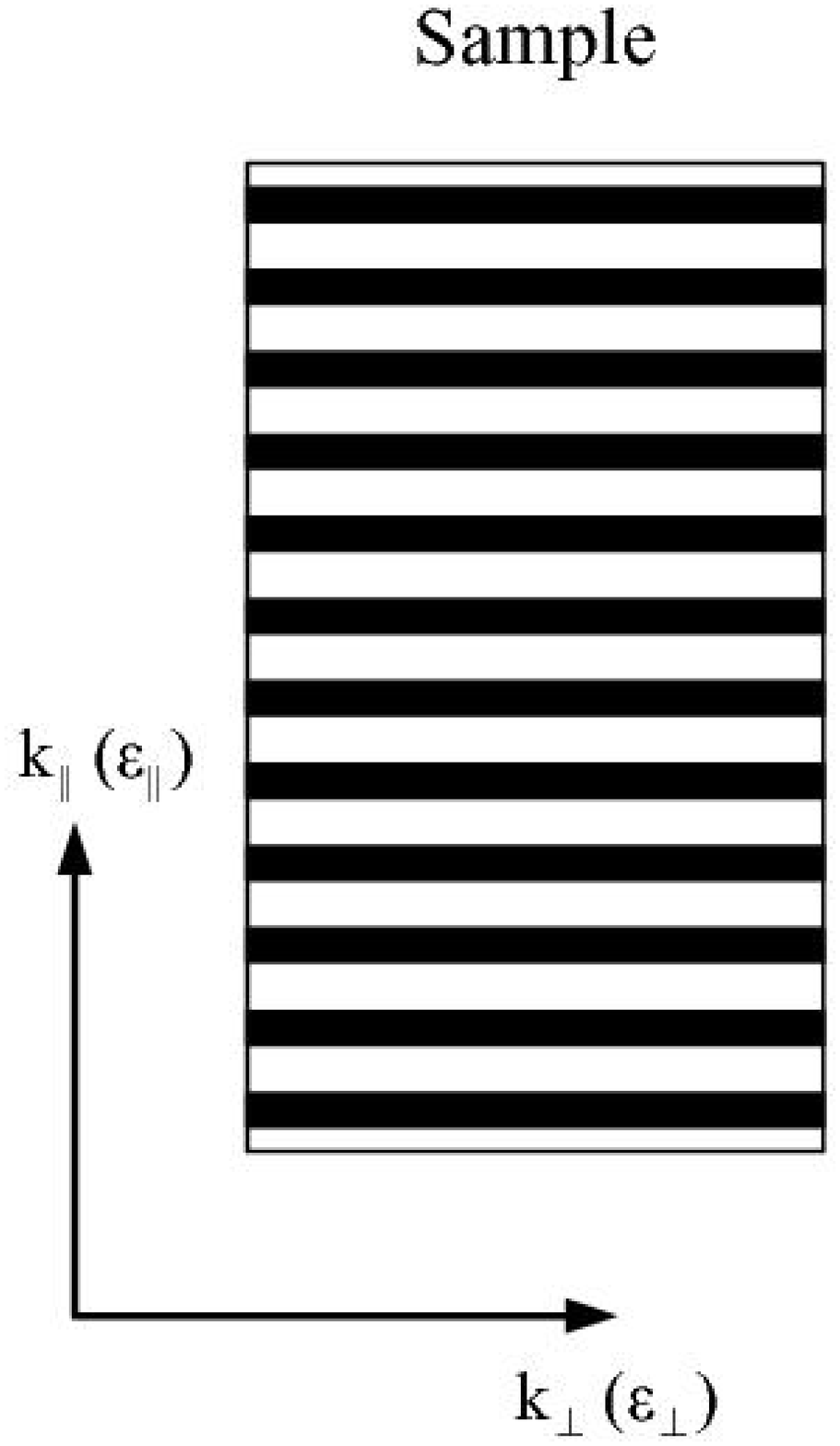}}
 \caption {%
  (Color online)
  (a) Top graph: Circular equifrequency surfaces (EFS) for vacuum and isotropic media. Bottom graph: Equifrequency surfaces for vacuum (circle) and negative anisotropic refraction media (hyperbolic
  relation). (b) The definitions for $k_{\perp}$, $k_{\parallel}$, $\varepsilon_{\perp}$, and $\varepsilon_{\parallel}$ used in our simulations.
 }%
 \label{schematic_illustration}
\end{figure}

For a particular anisotropic medium, where the permittivity component ($\varepsilon_{\perp}$) along the direction perpendicular to the interface is negative, while all other permittivity and permeability components are positive, it has a hyperbolic dispersion relation as follows:
\begin {equation}\label{dispersion}
 \frac
 {k_{\perp}^2}{|\varepsilon_{\parallel}|~\mu}-\frac
 {k_{\parallel}^2}{|\varepsilon_{\perp}|~\mu}= \omega^2
\end {equation}
where the definitions for $\varepsilon_{\perp}$, $\varepsilon_{\parallel}$, $k_{\perp}$ and $k_{\parallel}$ are shown in Fig.~\ref {schematic_illustration}(b). Figure~\ref {schematic_illustration}(a) schematically shows how negative refraction works in this particular anisotropic medium. The group velocity can be calculated by $\mathbf {v}_{g}=\nabla_{\mathbf {k}} \omega (\mathbf {k})$, which implies that the direction of group velocity (energy flow) would be normal to the equifrequency surface (EFS) and in the direction where $\omega$ is increasing. The conservation of $k_{\parallel}$ indicates two  possible solutions in the medium, but the correct one can be determined by causality --- the refracted group velocity should point away from the interface, as shown in Fig.~\ref {schematic_illustration}(a). From Fig.~\ref {schematic_illustration}(a), we can also see that for an isotropic medium, the circular equifrequency surface forces the refracted phase and group velocities to lie in the same line --- antiparallel for a negative index medium, while parallel for a positive index medium. For an anisotropic medium with a hyperbolic dispersion relation, they do not lie in the same line any more except for the case when $k_{\parallel}=0$. To be normal to the hyperbolic curves and satisfy the requirement of pointing away from the interface coming from the causality, the refracted group velocity has to undergo a negative refraction, which causes the expected focusing. (Note that the refracted phase velocity for an anisotropic medium still has a positive refraction.)

A lot of work has been done in anisotropic metamaterials, both experimentally \cite {hoffman07,yao08} and theoretically. \cite {smith03,smith04,Silveirinha07,wangberg06,lindell01,scalora07,wood06,jacob06,Liu08} Liu and Zhang \cite {Liu08} derived the hyperbolic dispersion only theoretically in the Maxwell-Garnett approximation. Although they showed negative refraction and pseudo focusing in numerical simulations, they did not obtain the actual dispersion relation from the realistic simulated metamaterial nor did they demonstrate the effective medium behavior from realistic simulations. There is a need to demonstrate that the hyperbolic dispersion survives all the way up to the evanescent waves, which is essential for potential super-resolution. Silveirinha et al., \cite {Silveirinha07} apart from analytical calculations, also did not demonstrate the hyperbolic dispersion of the simulated metamaterial. They only showed the near-field imaging (channeling), which occurs for the special case of a very flat dispersion. Yao et al. \cite {yao08} did experimental work (negative refraction for small angles only and no dispersion relation was obtained from the experiments), and Wangberg et al. \cite {wangberg06} presented analytical work based on the Maxwell-Garnett approximation. Most of the previous theoretical and numerical work on anisotropic metamaterials is done on homogeneous materials, where the hyperbolic dispersion relation given by Eq.~(\ref {dispersion}) is used. 

In this paper, we use realistic simulations for three-dimensional (3D) wire media and metal-dielectric superlattices to establish directly that the hyperbolic dispersion relation is valid up to evanescent modes in the long-wavelength limit and then retrieve the effective permittivity. A fitting procedure is exploited to get the dispersion relation from the field distributions obtained from full-wave numerical simulations of realistic structures. The imaging for a homogeneous slab with the effective permittivity shows very good agreement with the realistic structure. (All simulations about this homogeneous effective anisotropic medium are done by comsol multiphysics, an electromagnetic (EM) solver based on the finite element method.) We have three significant contributions to the field of anisotropic metamaterials: (1) the numerically obtained dispersion relations, (2) the demonstration of the effective medium behavior that works with evanescent incident modes and (3) our unique method to obtain the dispersion relations, different from the usual retrieval procedure based on inverting the scattering amplitudes.

\begin{figure}
  \centering
   \subfigure[]{
   \includegraphics[width=0.21\textwidth]{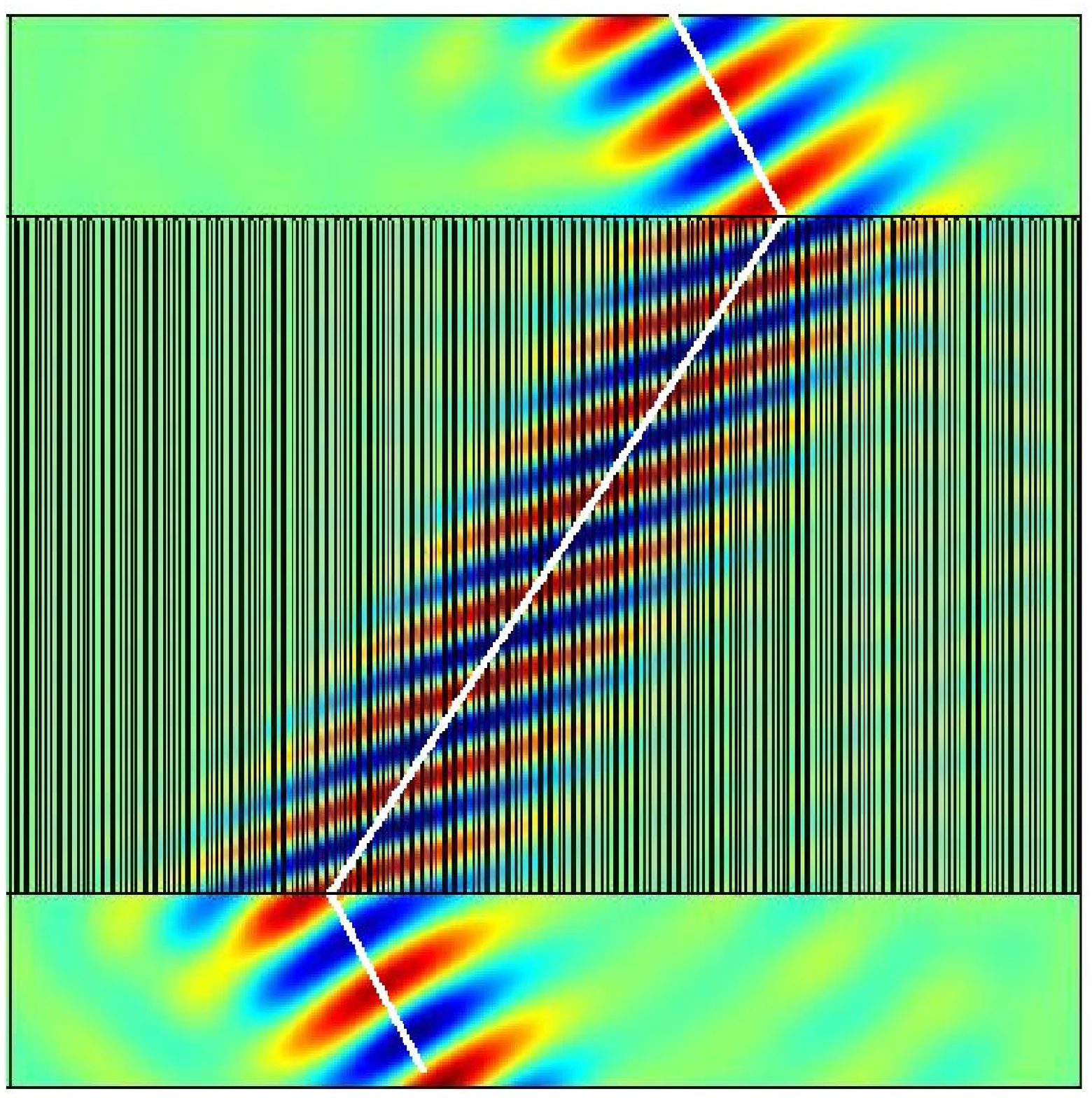}}
  \centering
   \subfigure[]{
   \includegraphics[width=0.21\textwidth]{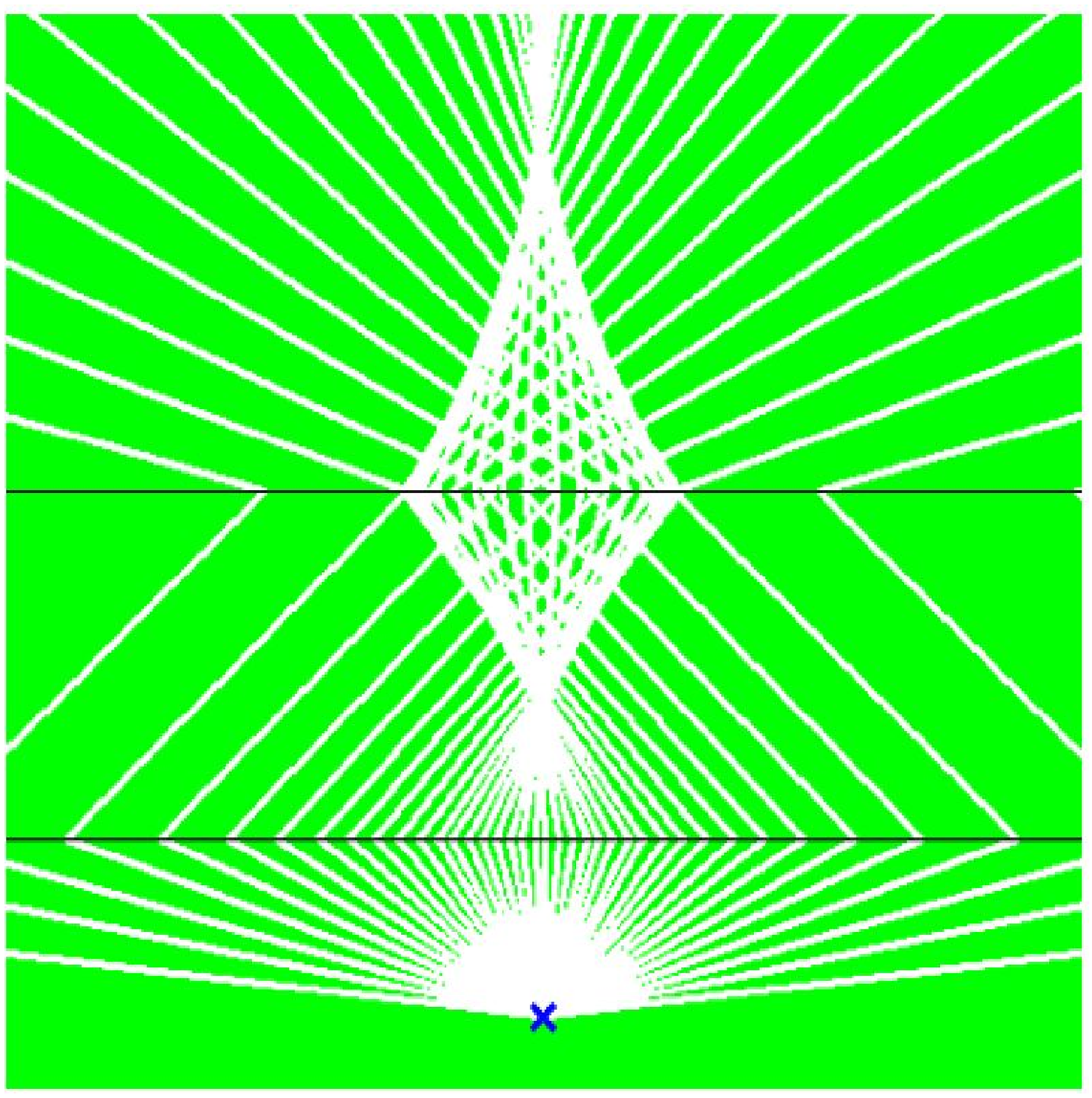}}
  \centering
   \subfigure[]{
   \includegraphics[width=0.21\textwidth]{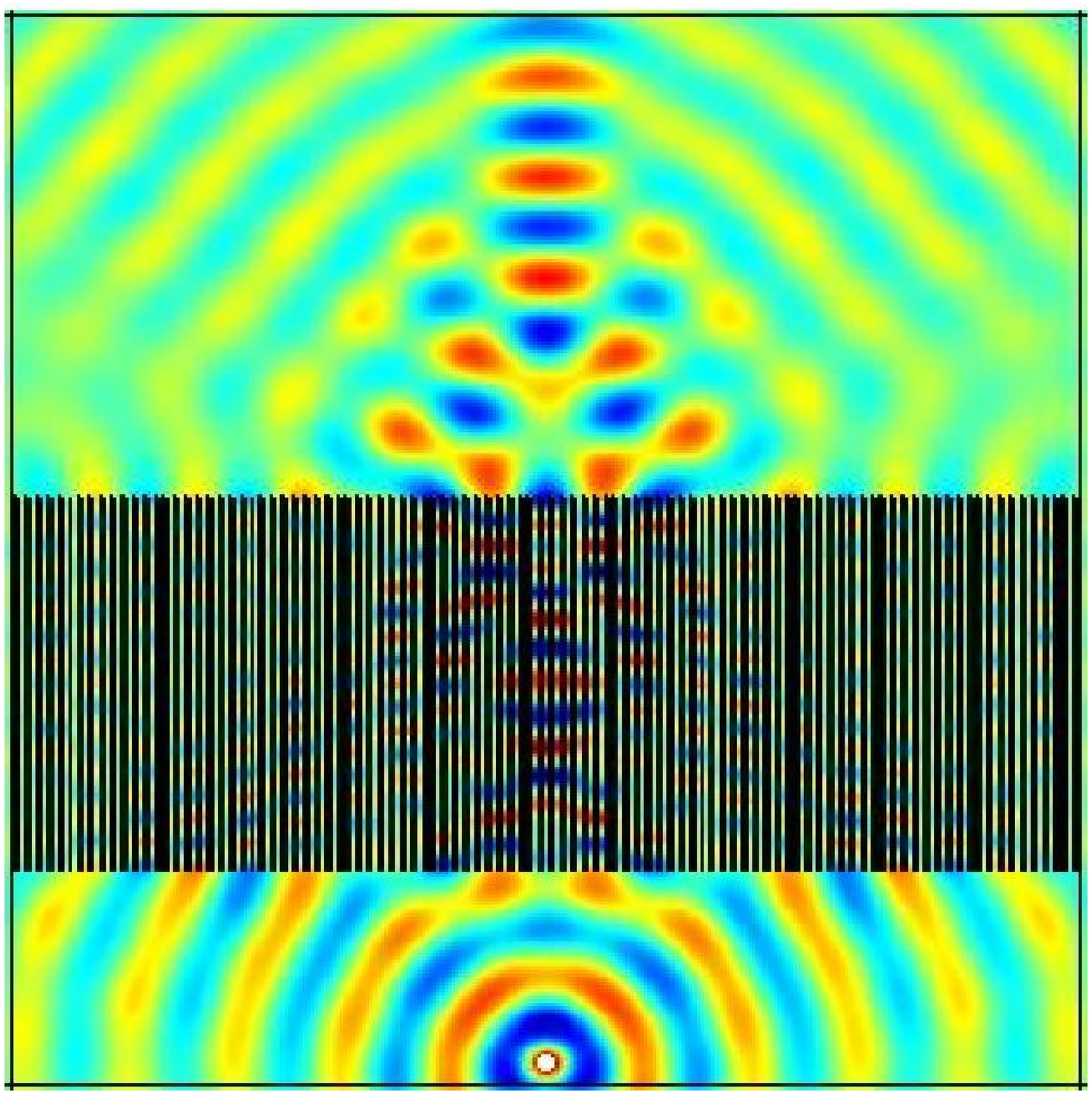}}
  \centering
   \subfigure[]{
   \includegraphics[width=0.21\textwidth]{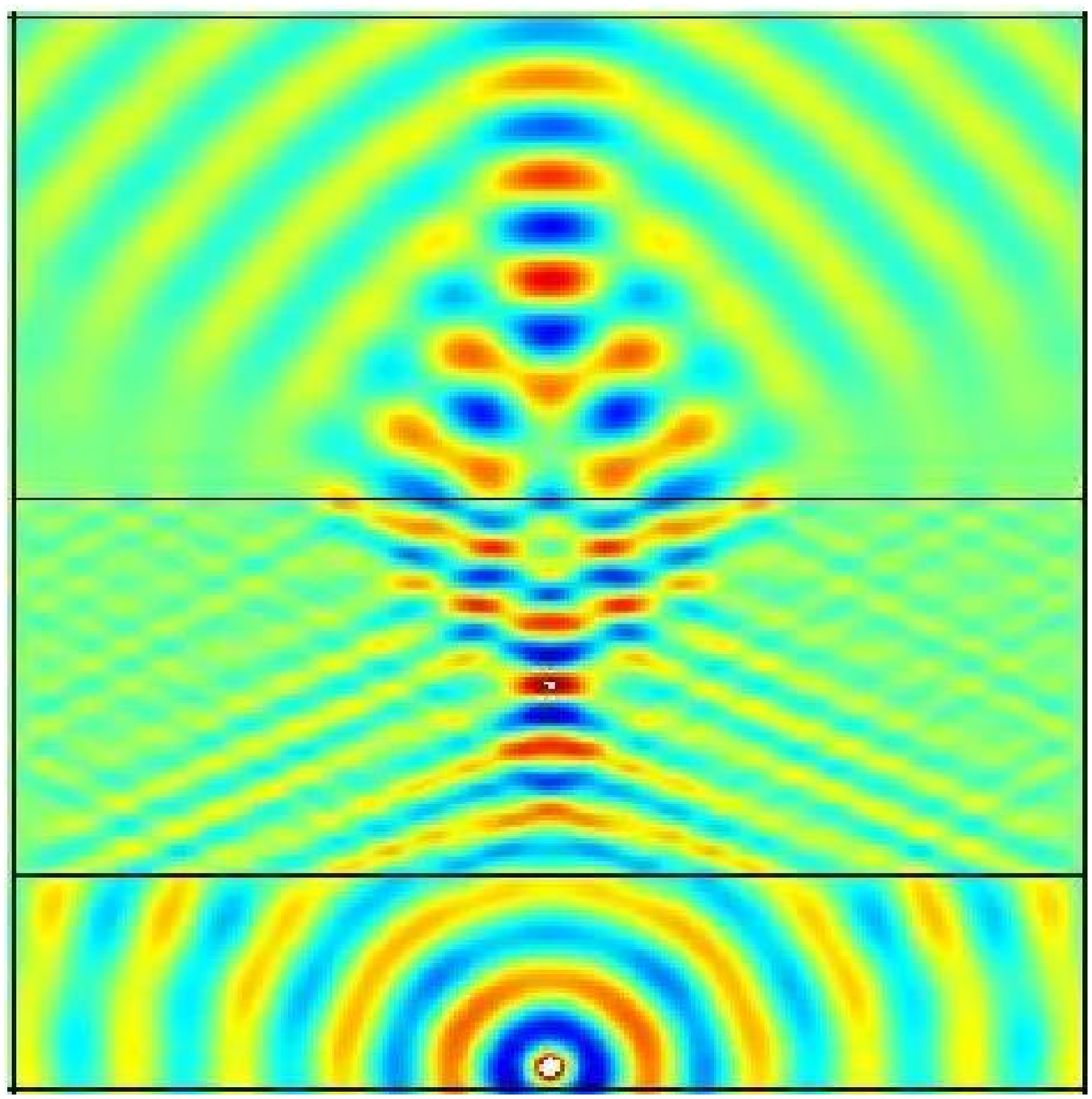}}
  \caption {%
  (Color online)
  Material parameters of the metallic layers: $\varepsilon = -4$, $\mu = 1$; working frequency $f=0.5~\textrm {GHz}$; space period of metallic layers is 0.06 m; width of metallic layers is 0.02 m. 
  (a) The magnetic field distribution of the group negative refraction in the metallic-air layers array slab. A Gaussian beam with the transverse magnetic polarization and an incident angle 
  of $30^0$ is incident on the simulated metallic-air superlattice. The white line indicates the ray-tracing result. (b) A ray-tracing diagram showing that the rays coming from a line source are 
  refocused by an anisotropic medium slab with the effective permittivity $\varepsilon$ tensor of our simulated metallic-air layers array slab. (c) The magnetic field distribution of the pseudo 
  focusing of our simulated metallic-plates array slab with a line source placed 1.25 m from the interface, which launches a cylindrical transverse magnetic polarized wave. The thickness and the 
  width of the metallic-air layers array slab are 2.4 m and 6 m, respectively. (d) The magnetic field distribution of the pseudo focusing in a homogeneous anisotropic slab with the effective 
  permittivity of our simulated metallic-air layers array slab ($\varepsilon_{\parallel}=1.7293$, $~\varepsilon_{\perp}=-0.7907$).
  }%
  \label {2Dfocusing}
\end{figure}

\section {Superlattice of metallic-air layers \label {superlattice}}
Before discussing our results on 3D wire media, simulations are performed for a superlattice of metallic layers with $\varepsilon=-4$ and air layers with $\varepsilon=1$ as shown in Fig.~\ref {schematic_illustration}(b). These simulations are done to check the applicability of our idea that one can obtain negative refraction and focusing in anisotropic media. Our simulation results show the existence of negative refraction in Fig.~\ref {2Dfocusing}(a) and pseudo-focusing in Fig.~\ref {2Dfocusing}(c). The focusing simulation is compared with the ray-tracing diagram (Fig.~\ref {2Dfocusing}(b)) and the imaging of a homogeneous anisotropic slab with the effective permittivity extracted from the dispersion relation of the metallic-air superlattice (Fig.~\ref {2Dfocusing}(d)). The effective parameters for $\varepsilon_{\parallel}$ and $\varepsilon_{\perp}$ are obtained by extracting $\mathbf {k}$ from the field distribution of a plane wave incidence inside the slab and then fitting with the hyperbolic dispersion given by Eq.~(\ref {dispersion}). The details for obtaining the effective parameters $\varepsilon_{\parallel}$ and $\varepsilon_{\perp}$ will be discussed below. One can see the pseudo focusing for the real metallic-air superlattice agrees very well with the homogeneous medium.

\section {Obtained numerical dispersion relations \label {numericalmethod}}
 To check if the wire medium constitutes our desired homogeneous effective anisotropic medium, it is straightforward to obtain its numerical dispersion relation first. For this purpose, we exploit a fitting procedure to extract $\mathbf k$ from the phase propagation. In the long wavelength limit, electromagnetic metamaterials should behave like a homogeneous medium. When a plane wave incidents on a homogeneous slab with an incident angle $\theta_i$, it forms a stationary wave inside the slab instead of a traveling wave because of the reflections at the two interfaces. Since $k_{\perp}$ represents the field variation in the perpendicular direction, we can take a cross-section along this direction and analytically obtain  the field distribution in the cross-section by considering the multireflections inside the slab as follows:
\begin{eqnarray}\label {kextraction}
  F(y) & = & \frac {A}{1-r^2\,e^{-2\,\alpha\,d}e^{2\,i\,k_{\perp}\,d}} \bigg [e^{-\alpha\,(y-y_0)}e^{i\,[k_{\perp}\,(y-y_0)+\theta]}  \nonumber \\
 & & +r\,e^{\alpha\,(y-y_0-2d)}e^{-i\,[k_{\perp}\,(y-y_0-2d)-\theta]}\bigg].
\end{eqnarray}
Here $y$ is the position in the perpendicular direction within the cross section, $F(y)$ is the field at the position $y$, $A$ and $\theta$ are the field amplitude and the field phase, respectively, at the starting point of the cross-section in the perpendicular direction $y=y_0$ (i.e., the location of the first interface of the slab), $\alpha$ is the decay factor of the homogeneous slab, $k_{\perp}$ is the perpendicular component of the wave vector $\mathbf {k}$, $d$ is the thickness of the slab, and $r$ is the reflection coefficient at the two interfaces.

By fitting the numerically obtained field distribution along the perpendicular direction in a cross-section with the theoretical formulas above, we can obtain the $k_{\perp}$ inside the wire medium slab for an incident plane wave with an incident angle, $\theta_i$. For $k_{\parallel}$, we can easily get $k_{\parallel}=k_0 \sin\theta_i$ from the incident angle $\theta_i$, since $k_{\parallel}$ is conserved across the interfaces, where $k_0$ is the wavevector in the background. Consequently, we can have the numerical dispersion relation of the wire medium by obtaining $k_{\parallel}$ and $k_{\perp}$ for different incident angles.

The minimum mean square fit does, in effect, average the field distribution on length scales small compared to the fitted effective wavelength. So the effective parameters are obtained for the averaged macroscopic field. The choice of the cross-section for the fit is arbitrary, but the results are practically independent on the location of the cross-section.

\section {3D anisotropic wire medium \label {3Dwires}}

\begin{figure}
   \subfigure[]{
   \includegraphics[ width=0.375\textwidth]{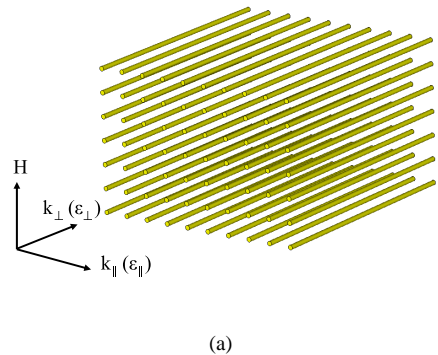}}
   \subfigure[]{
   \includegraphics[width=0.25\textwidth]{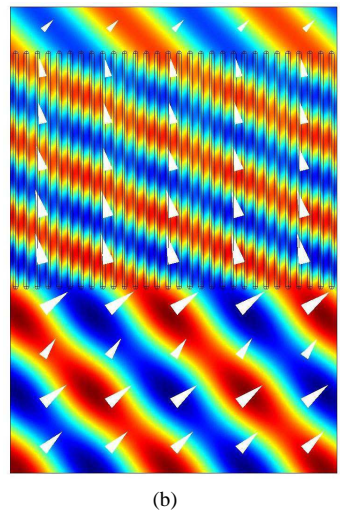}}
  \caption {%
  (Color online)
  (a) Schematic of 3D metallic wires embedded in a dielectric matrix. (b) The magnetic field distribution for the negative refraction in a 3D gold-wire square-lattice medium 
  with vacuum background and the wavelength $\lambda=700~\textrm {nm}$. The incident plane wave has transverse magnetic polarization and an incident angle of $45^0$. The 
  permittivity $\varepsilon$ for gold is taken from experimental data \cite {Palik99}: $\varepsilon = -15.5931+i~1.2734$ at $\lambda = 700~\textrm {nm}$. The radius, the length of gold wires, and 
  the lattice constant are $16$, $1532$ and $70~\mathrm {nm}$, respectively. The white arrow indicates the direction of power flow. }%
  \label {goldwire_negativerefraction}
\end{figure}

\begin{figure}
  \centering
  \includegraphics[width=0.25\textwidth]{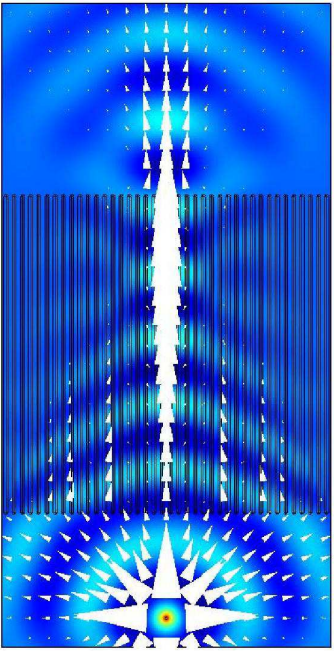}
  \caption {%
  (Color online)
  The magnetic field distribution of the pseudo focusing in a 3D gold-wire square-lattice medium  with a line source placed $884 ~\mathrm {nm}$ away from the interface, which launches a cylindrical 
  transverse magnetic polarized wave at the wavelength $\lambda=700 ~\textrm {nm}$. The permittivity of gold is the same as in  Fig.~\ref {goldwire_negativerefraction}. The background is vacuum. 
  The radius, length of gold wires, and the lattice constant are $16$, $2732$ and $70~ \mathrm {nm}$, respectively. The white arrow indicates the direction of power flow.
  }%
  \label {goldwire_focusing}
\end{figure}

The first structure for the 3D anisotropic wire medium in the optical region (Fig.~\ref {goldwire_negativerefraction} (a)) is a 3D gold-wire square lattice with the wire radius, $r=16~\textrm {nm}$, and the lattice constant, $a=70~\textrm {nm}$, in vacuum. Figure~\ref  {goldwire_negativerefraction} (b) shows that the group negative refraction occurs when a plane wave with the wavelength, $\lambda = 700 ~\textrm {nm}$, and the transverse magnetic polarization, incidents on our simulated slab with an incident angle of $45^0$, while the phase velocity still undergoes a positive refraction. Pseudo focusing can also be seen from Fig.~\ref {goldwire_focusing}, where the transverse magnetic polarized wave with the wavelength, $\lambda =700 ~\textrm {nm}$, coming out from a line source, is focused inside the simulated slab and then refocused on the other side of the slab.

\begin{figure}
  \centering
  \includegraphics[width=0.45\textwidth]{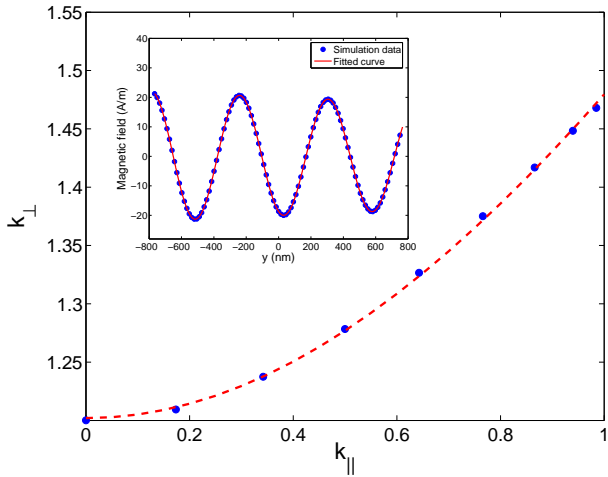}
  \caption {%
  (Color online)
  The numerical dispersion relation data from the simulation (solid circles) and the fitted hyperbolic curve (dashed line).  All parameters are the same as in  
  Fig.~\ref {goldwire_focusing}, except the length of gold wires which is $1500~ \textrm {nm}$. Note that all $k$ components here are normalized by $k_0$, where $k_0=\omega/c$. The 
  inset shows, as a typical example, the field distribution for $\theta_i = 30^0$ fitted by Eq.~(\ref {kextraction}).}%
  \label {goldwire_dispersion}
\end{figure}

When the geometric parameters, the wire radius $r=16~\textrm{nm}$, and the lattice constant $a = 70 ~\textrm {nm}$, are much smaller than the vacuum wavelength, $\lambda = 700~\textrm {nm}$, of the incident EM wave, the 3D wire medium can be considered as a homogeneous effective medium. \cite {elser06,sihvola99, foss94} The numerical dispersion relation of this 3D gold-wire square lattice medium is obtained and shown in Fig.~\ref {goldwire_dispersion}. The effective permittivities, $\varepsilon_{\perp}=-1.9082+i~0.2391$ and $\varepsilon_{\parallel}=1.4455+i~0.0044$, are obtained by fitting the numerical dispersion data into the hyperbolic dispersion relation (Eq.~(\ref {dispersion})). The fitted curve (dashed line) shows that the fitting is pretty good and the simulated metamaterial does have a hyperbolic dispersion relation.

\begin{figure}
  \centering
   \includegraphics[width=0.45\textwidth]{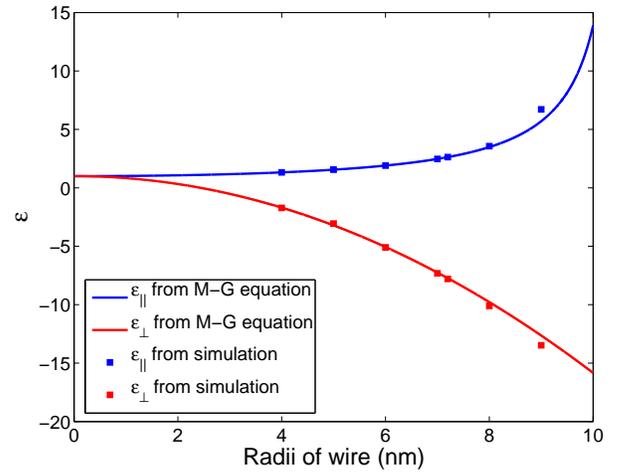}
  \caption {%
  (Color online)
  The effective permittivity $\varepsilon_{\perp}$ and $\varepsilon_{\parallel}$ calculated from Maxwell-Garnett equations (solid lines) and numerical simulations (squares) for different wire 
  radii. The simulated medium is a 3D square lattice silver wire medium in vacuum with the lattice constant $a = 20 ~\textrm {nm}$. The wavelength is $\lambda = 700 ~\textrm {nm}$. The 
  permittivity of silver at $\lambda= 700~\textrm {nm} $ is $\varepsilon_{\mathrm {silver}} = -20.4373+i~1.2863$, taken from experimental data. \cite {Palik99}}%
  \label {M_G_Equation}
\end{figure}

We have also used the Maxwell-Garnett equations \cite {sihvola99,foss94} to obtain the effective $\varepsilon_{\perp}$ and $\varepsilon_{\parallel} $ at $\lambda = 700 ~\textrm {nm}$ for different filling ratios for the square lattice of metallic wires. In Fig.~\ref {M_G_Equation}, we present the fitted results for $\varepsilon_{\parallel}$ and $\varepsilon_{\perp}$ for different radii, while keeping the lattice constant unchanged. We use the following expressions for $\varepsilon_{\parallel}$ and $\varepsilon_{\perp}$ from the Maxwell-Garnett theory:
\begin{equation}\label {MG1}
  \varepsilon_{\parallel} = \varepsilon_d \biggl [ \frac {(1+f)\varepsilon_m+(1-f)\varepsilon_d}{(1-f)\varepsilon_m+(1+f)\varepsilon_d}\biggr ],
\end{equation}
\begin{equation}\label {MG2}
  \varepsilon_{\perp} = f \varepsilon_m+(1-f)\varepsilon_d,
\end{equation}
where $f$ is the filling ratio of the metal, and $\varepsilon_m$ and $\varepsilon_d$ are the permittivities of metal and dielectric, respectively. Notice that the effective values of $\varepsilon_{\perp}$ and $\varepsilon_{\parallel} $ agree reasonably well with our fitting procedure. This is due to the effect that the vacuum wavelength, $\lambda = 700~\mathrm{nm}$, is much larger than the lattice constant and the radius of the metallic wires. In other cases, the effective parameters given by Eqs.~(\ref {MG1}) and (\ref {MG2}) do not agree with our fitting procedure.

\begin{figure*}
  \centering
   \subfigure[]{
   \includegraphics[width=0.15\textwidth]{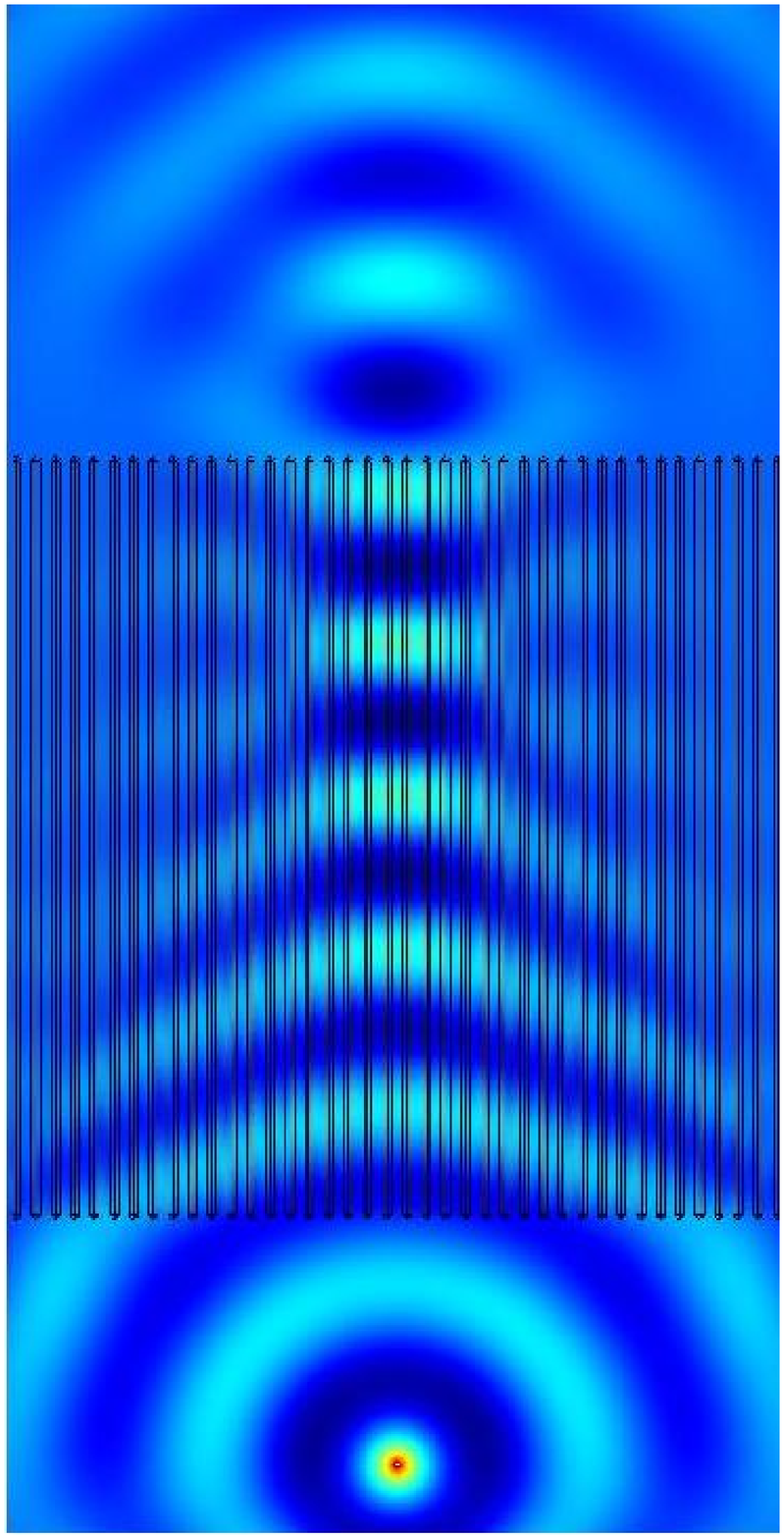}}
  \qquad
  \centering
   \subfigure[]{
   \includegraphics[width=0.15\textwidth]{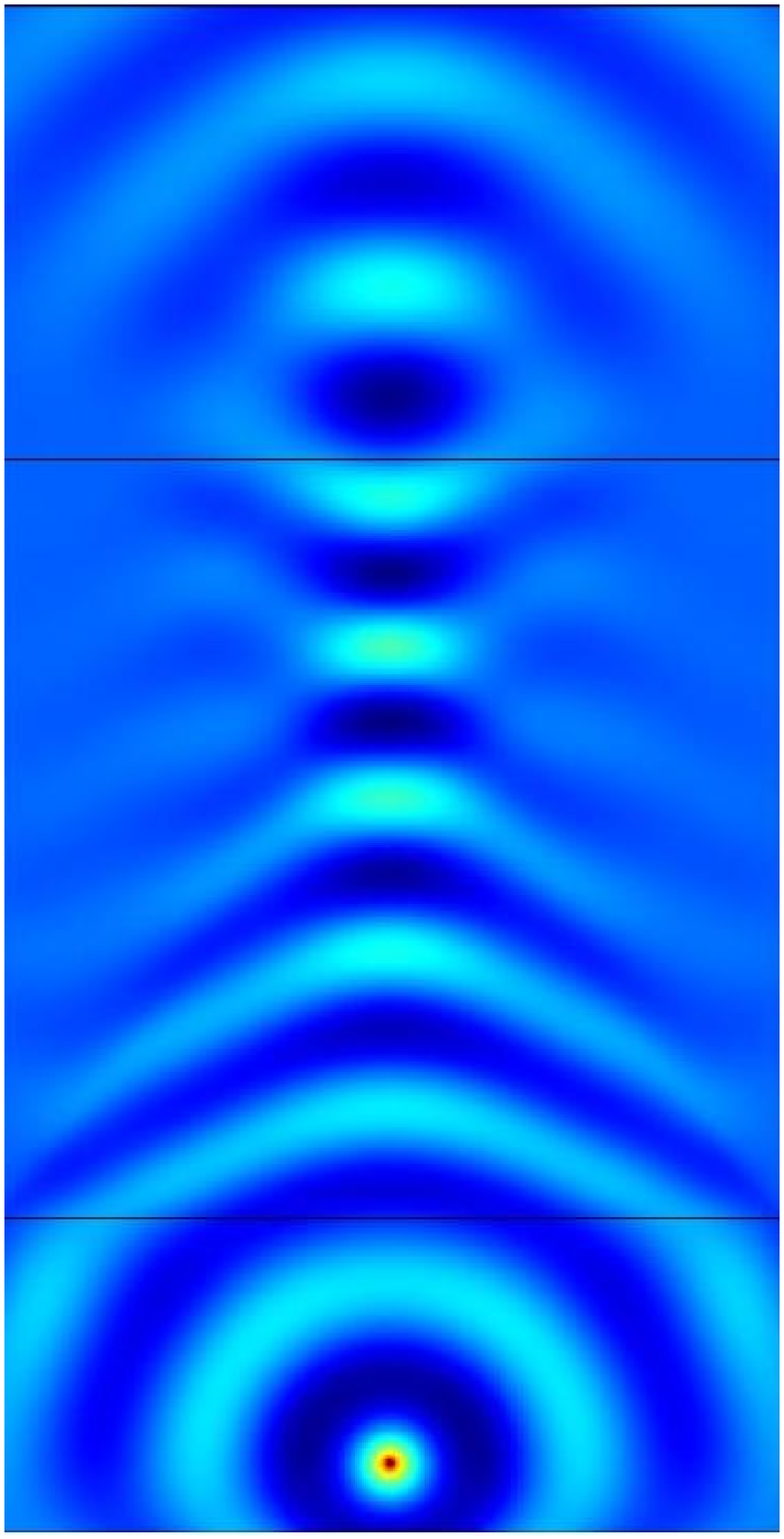}}
  \qquad
  \centering
   \subfigure[]{
   \includegraphics[width=0.15\textwidth]{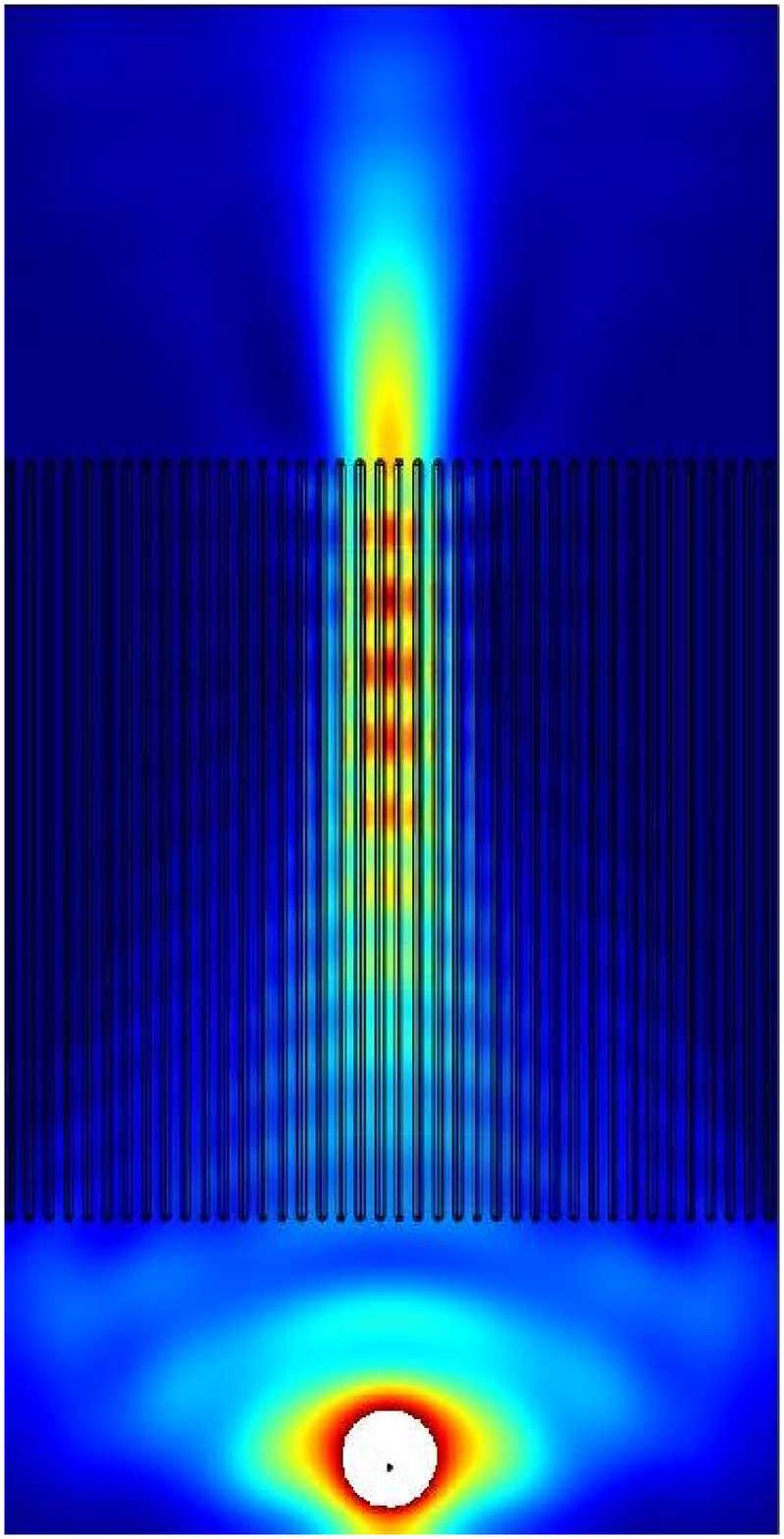}}
  \qquad
  \centering
   \subfigure[]{
   \includegraphics[width=0.15\textwidth]{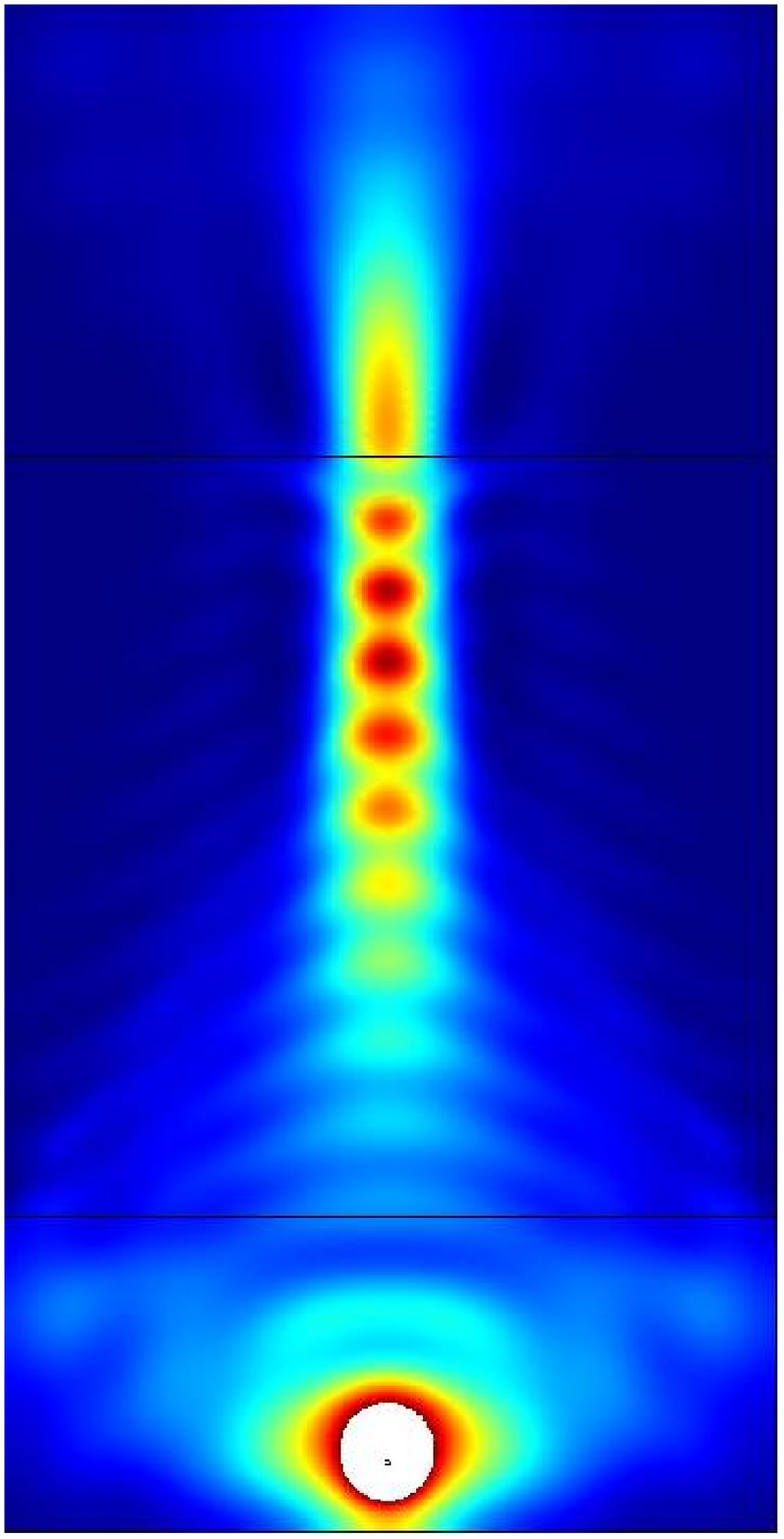}}
  \\
  \centering
   \subfigure[]{
   \includegraphics[width=0.15\textwidth]{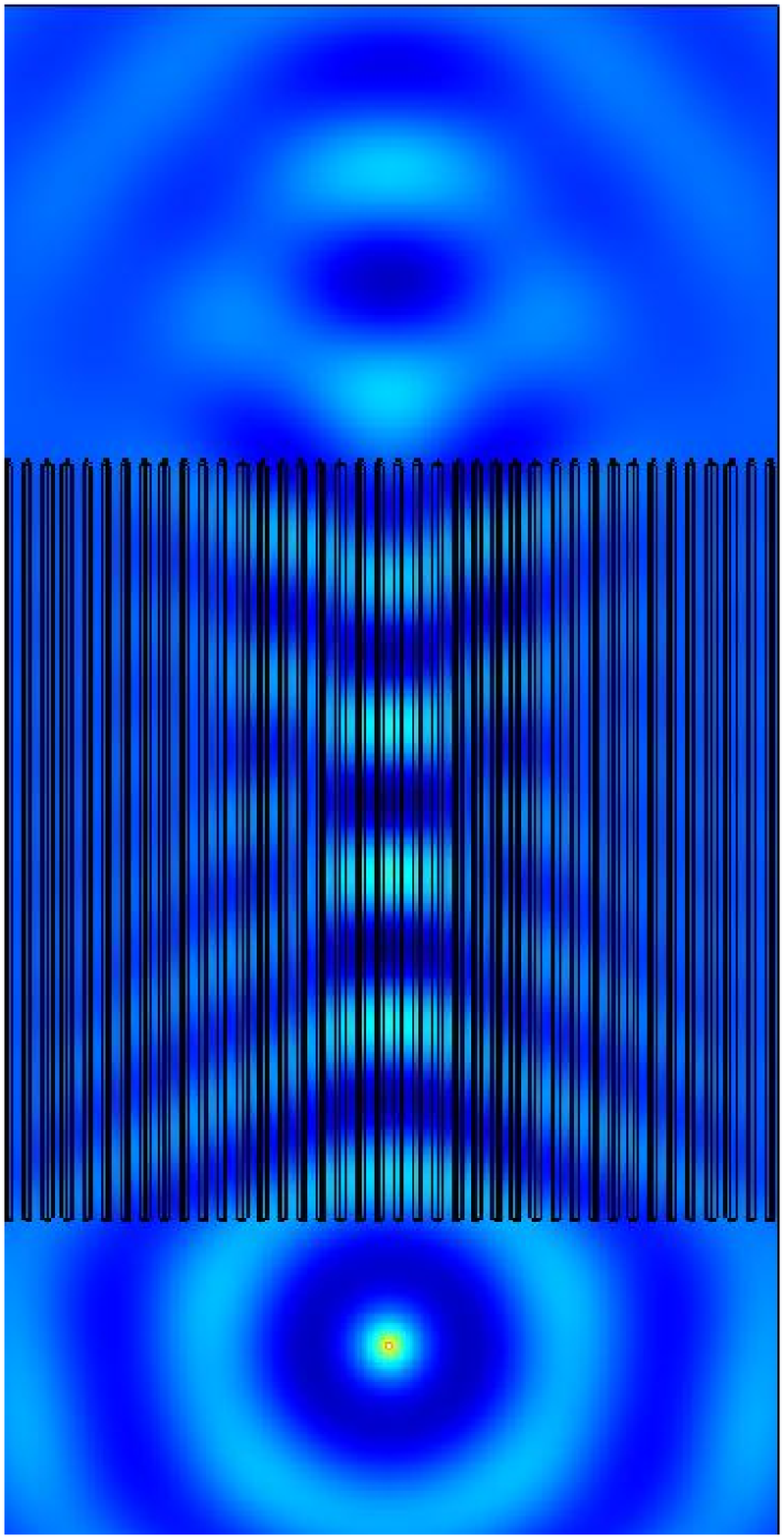}}
  \qquad
  \centering
   \subfigure[]{
   \includegraphics[width=0.15\textwidth]{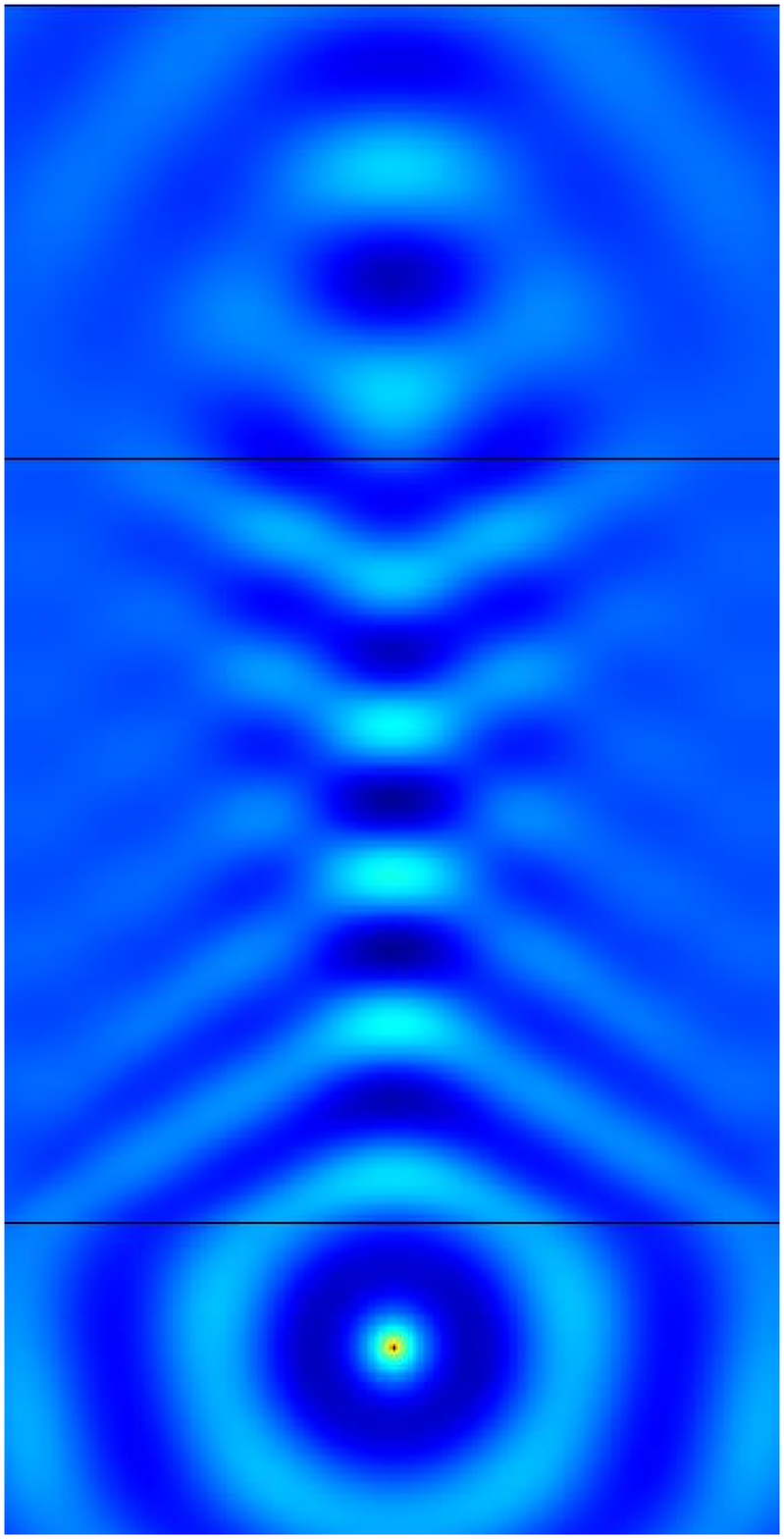}}
  \qquad
  \centering
   \subfigure[]{
   \includegraphics[width=0.15\textwidth]{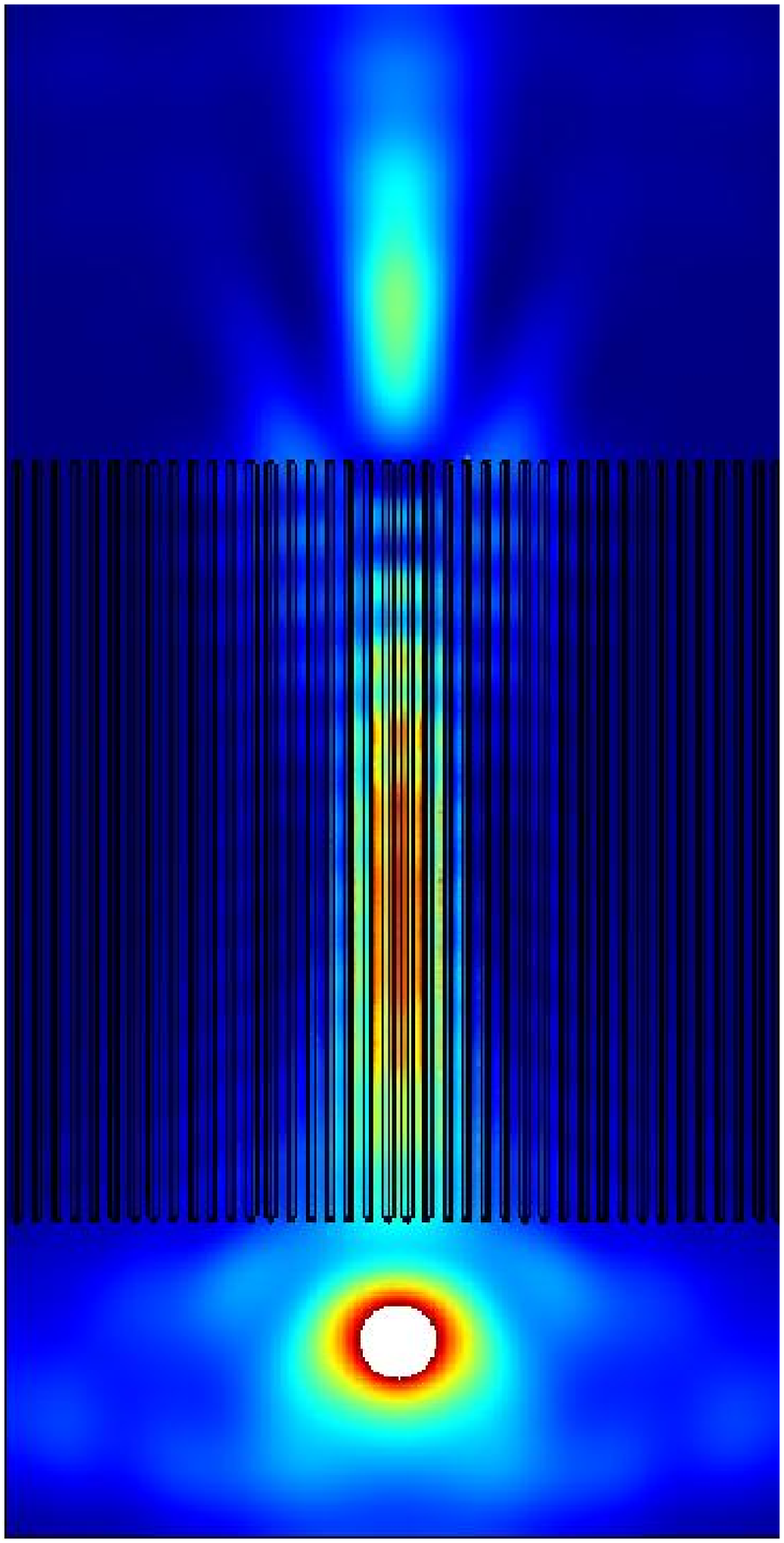}}
  \qquad
  \centering
   \subfigure[]{
   \includegraphics[width=0.15\textwidth]{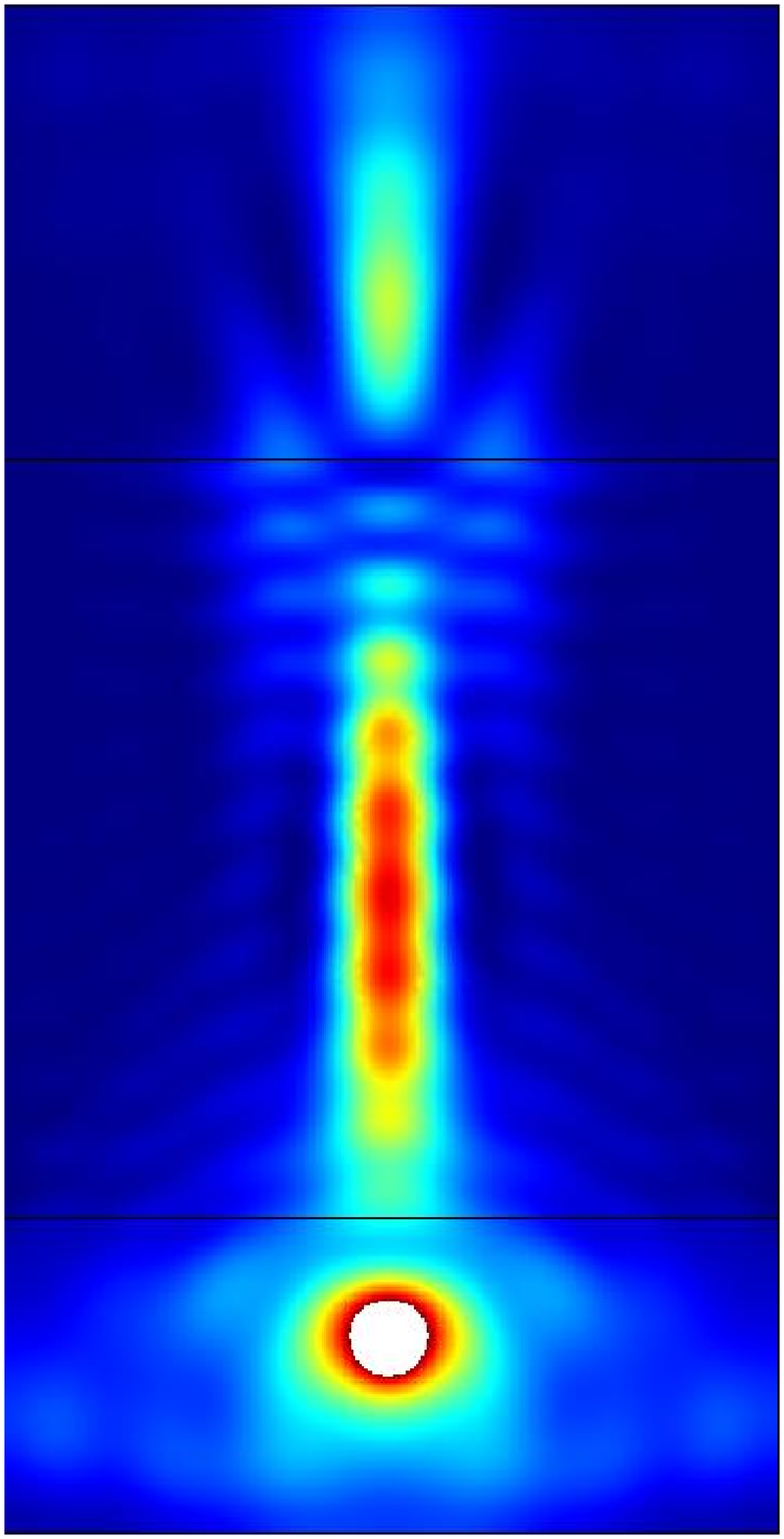}}
  \caption {%
  (Color online) 
  (a) The magnetic field distribution of the focusing simulation for the simulated 3D gold-wire square lattice anisotropic medium slab, with the source $884~\mathrm {nm}$ away from the first 
  interface. (b) Same as (a), but for a homogeneous anisotropic slab with 
  the fitted effective parameters $\varepsilon_{\parallel}=1.4455+i~0.0044$ and $\varepsilon_{\perp}=-1.9082+i~0.2391$. (c) and (d) are the same as (a) and (b), respectively, but for the magnetic 
  field intensity distribution. (e) -- (h) are the same as (a) -- (d), respectively, except the source is $442~\mathrm {nm}$ away from the first interface. All material parameters are the same as 
  in Fig.~\ref {goldwire_focusing}.}%
  \label {Goldwire_focusing_comparison}
\end{figure*}

For comparison, we replace this 3D gold-wire square lattice medium slab with a homogeneous anisotropic slab with  the fitted effective parameters $\varepsilon_{\parallel}=1.4455+i~0.0044$ and $\varepsilon_{\perp}=-1.9082+i~0.2391$. (All other parameters are the same, such as the thickness and the width of the slab, the source and the distance between the source and the first interface, etc.) The simulation results for the magnetic field distribution and magnetic field intensity are shown in Fig.~\ref {Goldwire_focusing_comparison}. One can see that both of them have very good agreements between the homogeneous slab and the 3D wire medium. The excellent agreement proves again that our simulated 3D gold-wire square-lattice metamaterial indeed behaves as an effective medium, which has a hyperbolic dispersion relation and our fitting procedure works very well.

\begin{figure}
  \centering
   \subfigure[]{
   \includegraphics[width=0.15\textwidth]{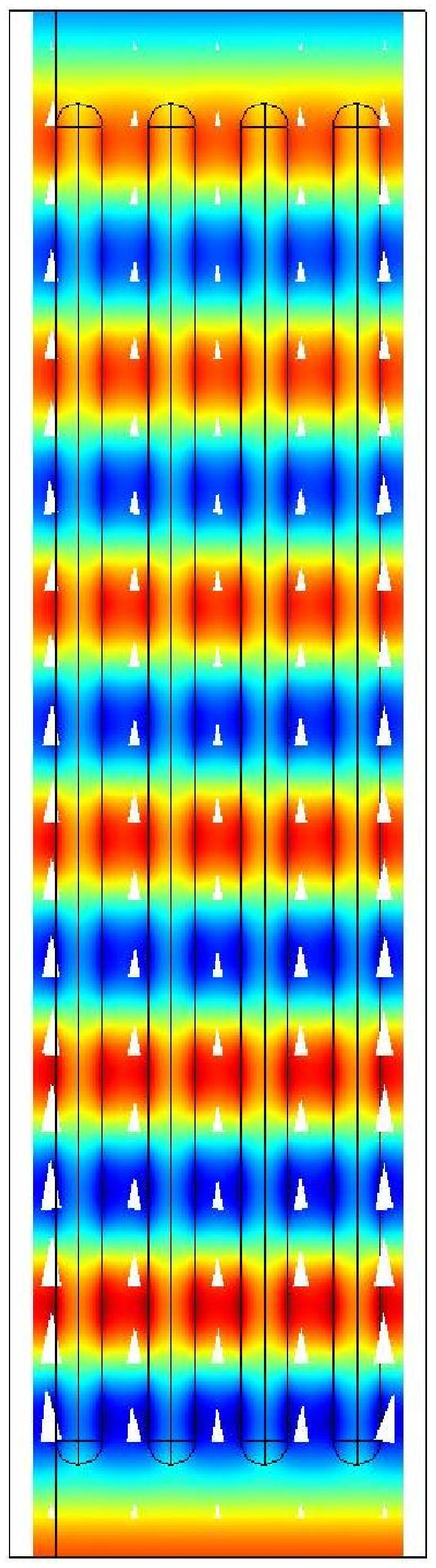}}
  \qquad
  \centering
   \subfigure[]{
   \includegraphics[width=0.15\textwidth]{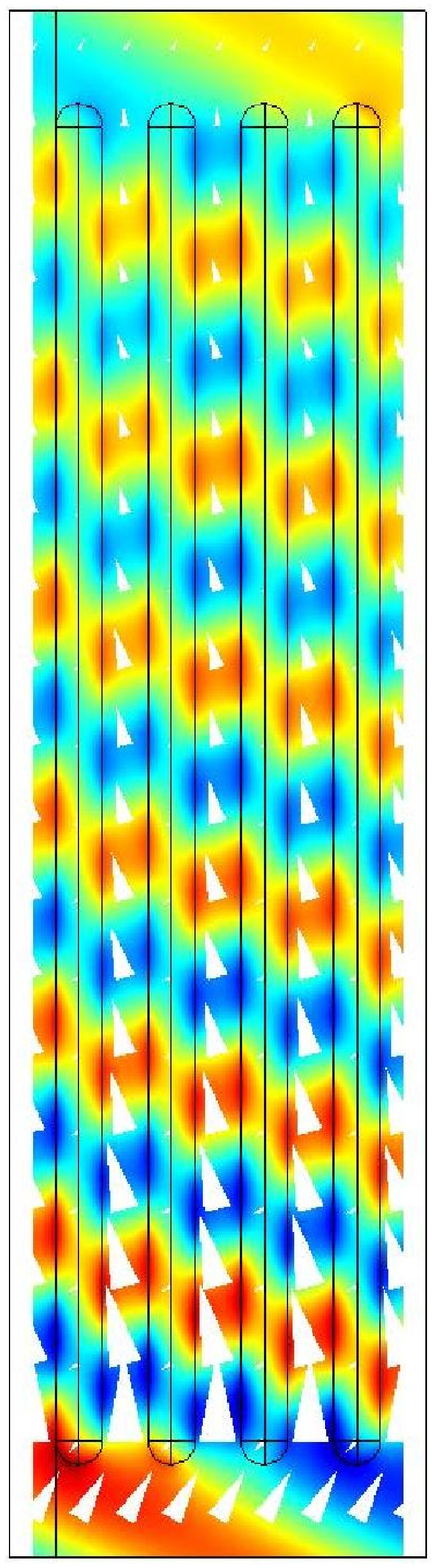}}
  \caption {%
  (Color online)
   The magnetic field distribution in a 3D silver-wire hexagonal lattice medium slab with the alumina background. The incident plane wave has the transverse magnetic polarization and the wavelength 
   in vacuum $\lambda = 700~\textrm {nm}$. (a) Normal incidence. (b) At an incident angle of $30^0$. The white arrow indicates the direction of power flow. The hexagonal lattice constant $a$, the 
   radius $r$, and the length $l$ of silver wires are $120~\textrm {nm}$, $30~\textrm {nm}$ and $1700~\textrm {nm}$, respectively. The permittivities of silver and alumina at the wavelength in vacuum 
  $\lambda = 700~\textrm {nm}$ are $\varepsilon_{\mathrm {silver}} =-20.4373+i~1.2863$ and $\varepsilon_{\mathrm {Al_{2}O_{3}}}=3.1$, respectively, taken from experimental data. \cite {Palik99} }%
  \label {Silver_Negative_Refraction}
\end{figure}

To be experimentally feasible, the second structure we examine is a hexagonal-lattice structure composed of silver wires in the alumina background. Figure~\ref {Silver_Negative_Refraction} shows the magnetic field distributions along a cross-section perpendicular to the magnetic field for two different incident angles ($0^0$ and $30^0$). For the incident angle $\theta$=$30^0$ case (Fig.~\ref {Silver_Negative_Refraction}(b)), one can see that the group velocity (white arrow) undergoes a negative refraction inside the simulated medium. A substantial decay in the perpendicular direction for the magnetic field and the power flow exists for both of these two different incident angles (Figs.~\ref {Silver_Negative_Refraction}(a) and \ref {Silver_Negative_Refraction}(b)), since the lossy metallic wires have a very high filling ratio in this particular wire medium.

\begin{figure}
  \centering
   \subfigure[]{
   \includegraphics[width=0.45\textwidth]{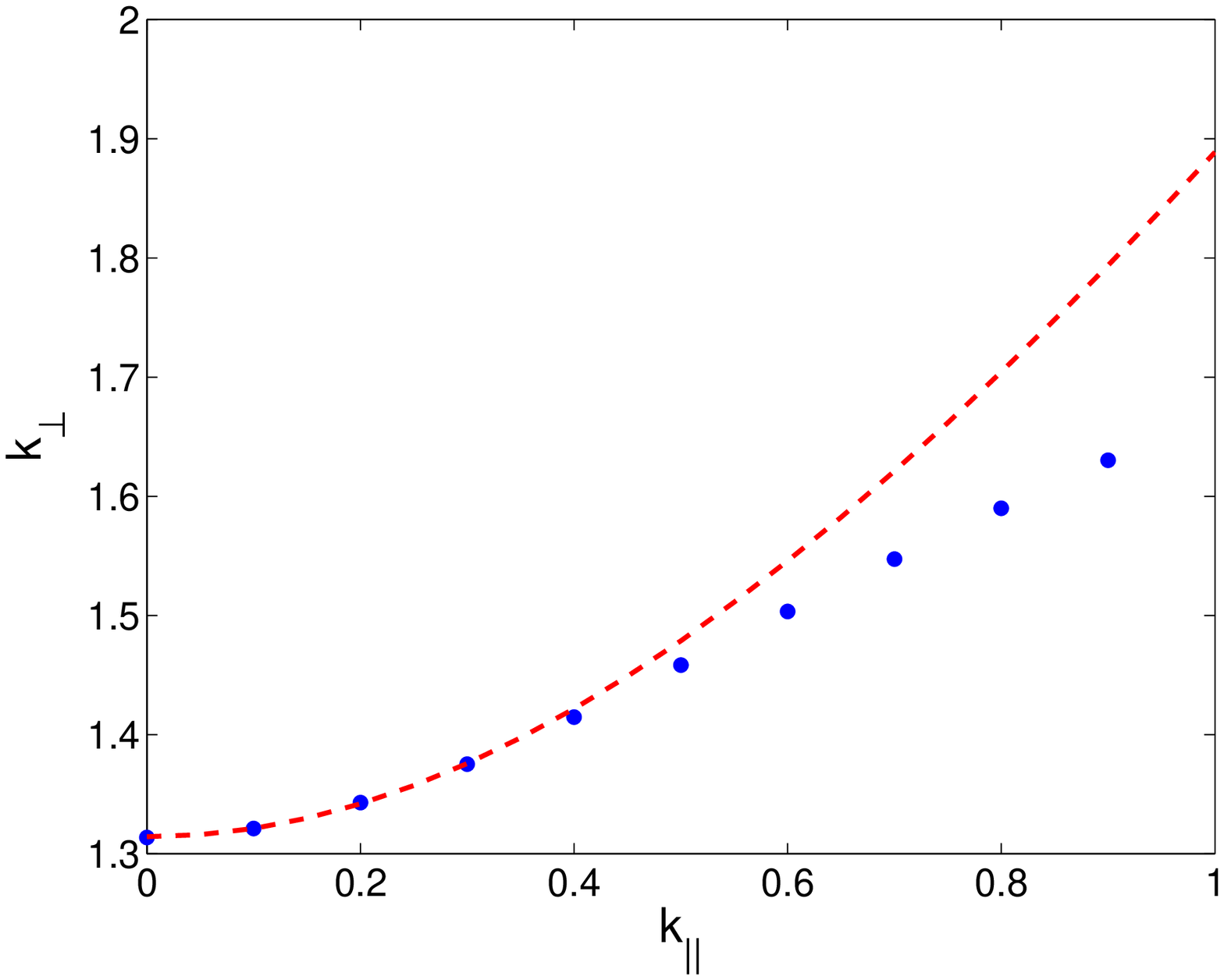}}
  \qquad
  \centering
   \subfigure[]{
   \includegraphics[width=0.45\textwidth]{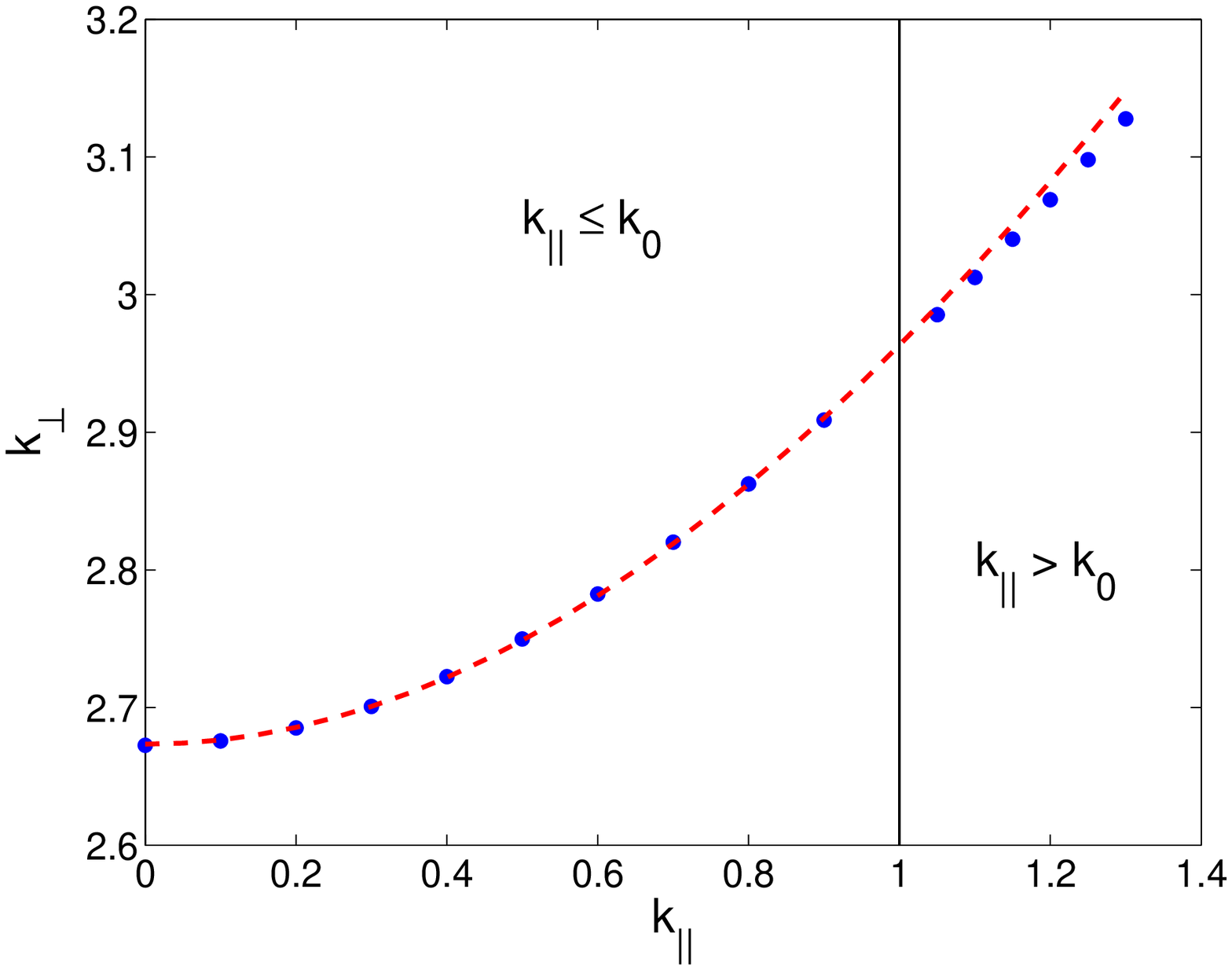}}
  \caption {%
  (Color online)
  The numerical dispersion (solid circles) and the fitted dispersion curve (dashed line) of 3D silver-wire hexagonal-lattice media in the alumina background. (a) The lattice 
  constant $a = 120~\textrm {nm}$ and the radius of silver wires $r = 30~\textrm {nm}$. (b) The lattice constant  $a = 30~\textrm {nm}$ and the radius of silver 
  wires $r = 12~\textrm {nm}$. $~k_{\parallel}\le k_0$ corresponds to the propagating modes in the background, while $k_{\parallel}>k_0$ corresponds to the evanescent modes. 
  All other parameters are the same as in Fig.~\ref {Silver_Negative_Refraction}. Note that all $k$ 
  components are normalized by $k_0$, where $k_0=\sqrt {\varepsilon}\omega/c$ and $\varepsilon$ is the permittivity of alumina.}%
  \label {Silver_Dispersion}
\end{figure}

By the same fitting procedure, the numerical dispersion relation for the 3D silver-wire hexagonal lattice  medium can also be obtained and is shown in Fig.~\ref{Silver_Dispersion}(a). The lowest four points are used to fit with a hyperbolic dispersion curve and the effective permittivity tensor is $\varepsilon_{\parallel}=5.3653+i~0.0708$ and $\varepsilon_{\perp}=-2.9188+i~0.4571$. One can see the large $k_{\parallel}$ points deviate from the fitted curve, even though the lowest four points are fitted very well. This occurs because we have a small wavelength/spatial period ratio of around 3.3 in alumina, which causes the breakdown of the homogeneous effective medium approximation in the large $k_{\parallel}$ region.

To extend the ``good fitted'' region to a larger $k_{\parallel}$ range, where the numerical dispersion points can fit well into a hyperbolic dispersion curve, we reduce the hexagonal-lattice constant and the radius of silver wires to smaller values $a=30~\textrm {nm}$ and $r=12~\textrm {nm}$, respectively, while keeping all other parameters the same as before, so we can have a much higher wavelength/spatial period ratio of around 13. The fitted numerical dispersion relation is shown in Fig.~\ref {Silver_Dispersion}(b). The lowest ten points, which are propagating modes (i.e., $k_{\parallel}\le k_0$, where $k_0=\sqrt {\varepsilon}\omega/c$ and $\varepsilon$ is the permittivity of alumina), are used to fit with a hyperbolic dispersion curve. The obtained effective permittivities are $\varepsilon_{\parallel}=22.1505+i~1.4693$ and $\varepsilon_{\perp}=-13.7714+i~0.6882$. If we use Eqs.~(\ref {MG1}) and (\ref {MG2}), the Maxwell-Garnett effective permittivities are given by $\varepsilon_{\parallel} = 25.8371 +i~ 2.0791$ and $\varepsilon_{\perp}=-10.5614 +i~0.7466$, which do not agree well with our fitting parameters. Maxwell-Garnet equations are an approximation, in particular, known to fail completely for the usual wire metamaterials in the microwave regime. Here, we include the comparison of the effective parameters derived directly from the simulated field distribution with those in the Maxwell-Garnet approximation to show that for high frequency (low permittivity) and ``thick'' wires the Maxwell-Garnet approximation becomes good and can be used to guide design. (The reason for this is the domination of the electron mass over the magnetic effective mass for the electrons geometrically confined to the wires at near optical length scales and frequencies.) In Fig.~\ref {Silver_Dispersion}(b), one can also see that the numerical dispersion relation data from our fitting procedure are fitted very well into a hyperbolic dispersion curve, even for those large $k_{\parallel}$ points, where $k_{\parallel}> k_0$. The latter are evanescent modes in air (and even in the alumina background of the wire medium), which are converted into propagating modes inside the slab and only attenuated by the losses of the effective medium. These modes preserve the information contained in the high spatial frequencies across the anisotropic slab and are essential for super-resolution applications.

\section{Conclusions}
We present two anisotropic metamaterials that demonstrate negative refraction and focusing. The first system is a superlattice of the metal-dielectric structure and the second system is (3D) metallic wires embedded in a dielectric matrix. We first obtain the numerical dispersion relation for the two cases by simulating the eigenmodes of the realistic system. The hyperbolic dispersion relation is obeyed in both cases, where the effective permittivities have opposite signs in the two propagation directions. Our simulations of the realistic structures, as well as the homogeneous simulations, show the negative refraction for all incident angles and demonstrate the focusing. The metallic nanowires can be valid for the evanescent modes in the dielectric background by having a large wavelength/spatial period ratio, which has important applications in super-resolution.

In conclusion, we numerically demonstrate that a homogeneous effective indefinite anisotropic medium can be realized by a 3D nanowire medium at the optical frequency region, which can have a negative refraction and pseudo focusing. We also present a nice fitting procedure by which we can obtain the numerical dispersion relation of our 3D wire medium and then retrieve its effective permittivity tensor. Meanwhile, we demonstrate that the hyperbolic dispersion relation of the 3D nanowire medium can be valid for the evanescent modes in the background by having a large wavelength/spatial period ratio (i.e., in the long wavelength limit), which may have important applications in super-resolution.

\appendix*

\begin{acknowledgments}
Work at Ames Laboratory was supported by the Department of Energy (Basic Energy Science) under Contract No. DE-ACD2-07CH11358. This work was partially supported by AFOSR under MURI grant (Grant No. FA 9550-06-1-0337) and the European Community project ENSEMBLE.
\end{acknowledgments}

\bibliographystyle{apsrev}

\end{document}